\documentclass[11pt]{article}
\usepackage{graphicx}

\newcommand{\BABARPubYear}    {02}

\newcommand{\BABARConfNumber} {17}
\newcommand{\SLACPubNumber} {9324}

\input pubboard/babarsym

\long\def\inst#1{\par\nobreak\kern 4pt\nobreak
    {\it #1}\par\vskip 10pt plus 3pt minus 3pt}

\def\PLB{{Phys. Lett.} B\,}

\def\PRD{{Phys. Rev.} D}
\def\PRL{{Phys. Rev. Lett.}}

\def\etal{{\it et al.}}

\providecommand{\thetathr}{\mbox{$\theta_{\rm thr}$}}
\providecommand{\thetasph}{\mbox{$\theta_{\rm sph}$}}

\def\stat{{\rm stat.}}
\def\syst{{\rm syst.}}

\providecommand{\mm}{\mbox{\rm mm}}

\providecommand{\MeV}{\mbox{\rm MeV}}
\providecommand{\MeVc}{\mbox{${\rm MeV}/c$}}
\providecommand{\MeVcsq}{\mbox{${\rm MeV}/c^2$}}

\providecommand{\GeVcsq}{\mbox{${\rm GeV}/c^2$}}

\providecommand{\ra}{\mbox{$\rightarrow$}}

\providecommand{\chisq}{\mbox{$\chi^2$}}

\providecommand{\nue}{\mbox{$\nu_e}$}

\providecommand{\epl}{\mbox{$e^+$}}
\providecommand{\emi}{\mbox{$e^-$}}
\providecommand{\eplemi}{\mbox{\epl \emi}}

\providecommand{\hzero}{\mbox{$h^0$}}
\providecommand{\hmi}{\mbox{$h^-$}}

\providecommand{\Wmi}{\mbox{$W^-$}}

\providecommand{\pipl}{\mbox{$\pi^+$}}
\providecommand{\pimi}{\mbox{$\pi^-$}}
\providecommand{\pizero}{\mbox{$\pi^0$}}

\providecommand{\twopi}{\mbox{\pipl \pimi}}
\providecommand{\threepi}{\mbox{\pipl \pizero \pimi}}

\providecommand{\rhomi}{\mbox{$\rho^-$}}

\providecommand{\qbar}{\mbox{$\overline{q}$}}
\providecommand{\ubar}{\mbox{$\overline{u}$}}
\providecommand{\dbar}{\mbox{$\overline{d}$}}
\providecommand{\cbar}{\mbox{$\overline{c}$}}

\providecommand{\bbar}{\mbox{$\overline{b}$}}

\providecommand{\Kbar}{\mbox{$\overline{K}$}}

\providecommand{\Kmi}{\mbox{$K^-$}}

\providecommand{\KL}{\mbox{$K^0_L$}}
\providecommand{\KS}{\mbox{$K^0_S$}}

\providecommand{\Kbar}{\mbox{$\overline{K}$}}

\providecommand{\Kstar}{\mbox{$K^*(892)$}}

\providecommand{\bbar}{\mbox{\bbar}} 
\providecommand{\cbar}{\mbox{\cbar}}
\providecommand{\qbar}{\mbox{\qbar}}

\providecommand{\degrees}{\mbox{$^\circ$}}

\providecommand{\Dbar}{\mbox{$\overline{D}$}}
\providecommand{\Dzero}{\mbox{$D^0$}}

\providecommand{\Dpl}{\mbox{$D^+$}}

\providecommand{\Dstar}{\mbox{$D^{*}$}}

\providecommand{\Dstarzero}{\mbox{$D^{*0}$}}

\providecommand{\Dstze}{\mbox{$D^{(*)0}$}}

\providecommand{\Bbar}{\mbox{$\overline{B}$}}
\providecommand{\BBbar}{\mbox{$B \overline{B}$}}                                                   
\providecommand{\Bzero}{\mbox{$B^0$}}
\providecommand{\Bzerobar}{\mbox{$\overline{B^0}$}}
\providecommand{\Bpl}{\mbox{$B^+$}}
\providecommand{\Bmi}{\mbox{$B^-$}}

\newcommand{\mc}{{\rm Monte Carlo}}
\def\Kpi{\mbox{$\Kmi\pipl$}}
\def\Ktwopi{\mbox{$\Kmi\pipl\pizero$}}
\def\Kthreepi{\mbox{$\Kmi\pipl\pipl\pimi$}}

\def\Bzerobar{\mbox{$\Bbar^0$}}

\def\mes{\mbox{$m_{\rm ES}$}}

\begin{document}
{\pagestyle{empty}

\begin{flushright}
\babar-CONF-\BABARPubYear/\BABARConfNumber \\
SLAC-PUB-\SLACPubNumber \\
July 2002 \\
\end{flushright}

\par\vskip 5cm

\begin{center}
\Large \bf \large Measurement of Branching Fractions of
Color-Suppressed Decays \\
of the $\mathbf{\Bzerobar}$ Meson to $\mathbf{\Dzero \pizero}$, 
$\mathbf{\Dzero \eta}$, and $\mathbf{\Dzero \omega}$ \\
\end{center}
\bigskip

\begin{center}
\large The \babar\ Collaboration\\
\mbox{ }\\
July 24, 2002
\end{center}
\bigskip \bigskip

\begin{center}
\large \bf Abstract
\end{center}
We report preliminary results of 
an experimental investigation of the color-suppressed decays
$\Bzerobar \ra \Dzero \pizero, \Dzero \eta$, and $\Dzero \omega$.
We measure the branching fractions 
${\cal B} (\Bzerobar \ra \Dzero \pizero) = 
(2.89 \pm 0.29 (\stat) \pm 0.38 (\syst)) \times 10^{-4}$,
${\cal B} (\Bzerobar \ra \Dzero \eta) = 
(2.41 \pm 0.39 (\stat) \pm 0.32 (\syst)) \times 10^{-4}$, 
and ${\cal B} (\Bzerobar \ra \Dzero \omega) = 
(2.48 \pm 0.40 (\stat) \pm 0.32 (\syst)) \times 10^{-4}$. 
The results are based on $(48.8\pm 0.5) \times 10^6$ $\BBbar$ pairs
collected with the \babar\ detector.
The branching fractions of these color-suppressed decays are 
significantly larger than theoretical expectations based upon 
factorization.

\vfill
\begin{center}
Contributed to the 31$^{st}$ International Conference on High Energy Physics,\\ 
7/24---7/31/2002, Amsterdam, The Netherlands
\end{center}

\vspace{1.0cm}
\begin{center}
{\em Stanford Linear Accelerator Center, Stanford University, 
Stanford, CA 94309} \\ \vspace{0.1cm}\hrule\vspace{0.1cm}
Work supported in part by Department of Energy contract DE-AC03-76SF00515.
\end{center}

\newpage
} 

\begin{center}
\small

The \babar\ Collaboration,
\bigskip

B.~Aubert,
D.~Boutigny,
J.-M.~Gaillard,
A.~Hicheur,
Y.~Karyotakis,
J.~P.~Lees,
P.~Robbe,
V.~Tisserand,
A.~Zghiche
\inst{Laboratoire de Physique des Particules, F-74941 Annecy-le-Vieux, France }
A.~Palano,
A.~Pompili
\inst{Universit\`a di Bari, Dipartimento di Fisica and INFN, I-70126 Bari, Italy }
% ...

\end{center}\newpage

\section{Introduction}
\label{sec:Introduction}

Decays of the type $\Bzerobar \ra \Dpl \hmi$ proceed through a color-allowed
spectator diagram in which the $\Wmi$ decays to a $\ubar d$ quark pair that
hadronizes into the light hadron $\hmi$.
On the other hand, decays of the type $\Bzerobar \ra \Dzero \hzero$ proceed
through a color-suppressed spectator diagram that requires
color matching of the quark and the antiquark from the $\Wmi$ with the $c$
and $\dbar$ quarks.
Since perturbative calculations of hadronic $B$ decay rates are
not possible, we must rely on models for predictions.
The naive factorization model predicts
very low branching fractions for color-suppressed decays, in the range
$0.3$--$0.7\times 10^{-4}$ \cite{ref:NeuSte,ref:Deandrea}.
However, this model is supported by HQET only for
color-allowed decays, while color-suppressed decays
receive substantial corrections \cite{ref:NeuPet} that depend upon
the decay mode.
Experimental measurements of the branching fractions of color-suppressed
$B$ decays provide an important test of theoretical models and can be
used to improve the models.

In this paper we report on the observation of the three color-suppressed
decays $\Bzerobar \ra \Dzero \pizero$, $\Bzerobar \ra \Dzero \eta$,
and $\Bzerobar \ra \Dzero \omega$.
The $\Bzerobar$ decay into $\Dzero \pizero$ has been
observed previously by the CLEO Collaboration \cite{ref:CLEO}, 
all three decays have been observed by the 
Belle Collaboration \cite{ref:Belle}.

\section{The \babar\ detector and dataset}
\label{sec:babar}
The data used in this analysis were collected with the \babar\ detector
at the \pep2\ asymmetric $\eplemi$
storage ring at the $\Upsilon(4S)$ resonance, 
between October 1999 and December 2001.
This data sample contains $(48.8 \pm 0.5) \times 10^6$ $\Bzero\Bzerobar$ and 
$\Bpl\Bmi$ pairs.

The \babar\ detector is described in detail elsewhere \cite{ref:babar}.
We briefly summarize the detector systems most relevant to 
this analysis.
The \babar\ detector contains a 5-layer silicon vertex tracker (SVT)
and a 40-layer drift chamber (DCH) situated in a 1.5\,T solenoidal
magnetic field.  
These devices detect charged particles and measure their momentum
and ionization energy loss ($dE/dx$).
Surrounding the DCH are fused-silica quartz bars of a 
ring-imaging Cherenkov detector (DIRC).
This detector measures the Cherenkov angle of light 
generated in the bars.
The charged particle identification (PID) 
combines the $dE/dx$ measurements of the SVT, DCH, and DIRC.
Photons are detected in a CsI(Tl) crystal electromagnetic calorimeter (EMC).
The EMC detects photons with energies as low as 20\,\MeV.

The interactions of particles traversing the detector are simulated using 
the GEANT4 \cite{ref:GEANT} program.
Beam-induced backgrounds are taken into account.
Signal and generic background \mc\ samples are used to study the effect of
the event selection criteria and to estimate the backgrounds.
The generic background \mc\ simulation consists of 
$\eplemi \ra \qqbar$ $(q=u,d,s,c)$
and $\Bpl\Bmi$ and $\Bzero\Bzerobar$ events.

\section{Analysis method}
\label{sec:Analysis}

Here, we describe the reconstruction and the selection of the three 
color-suppressed modes 
$\Bzerobar \ra \Dzero \pizero$, $\Bzerobar \ra \Dzero \eta$, 
and $\Bzerobar \ra \Dzero \omega$.

\subsection{Particle selection}
\label{subsec:partsel}

Photons are identified by energy deposits in contiguous crystals in the EMC.
They must have an energy greater than $30\,\MeV$ and a 
lateral shower shape compatible with electromagnetic showers.
Charged particle tracks (except those used to reconstruct $\rho$ mesons) 
must have at least 12 hits in the DCH and $p_t > 100\,\MeVc$.
Tracks must extrapolate to within $20\,\mm$ of the $\eplemi$ interaction 
point in the plane transverse to the beam axis 
and within $50\,\mm$ along the beam axis.
Charged kaon candidates are identified using a likelihood function that 
combines $dE/dx$ and DIRC information.
The likelihood function is used to define a tight kaon criterion 
and a loose criterion to veto pions.
To satisfy the tight kaon criteria, the track must also have 
$p_t > 250\,\MeVc$ and an angle with the beam between 0.45 and 2.5 rad 
so that the candidate is within the fiducial region of the DIRC.

\subsection{Light hadron and $D$ meson reconstruction}
\label{subsec:X0andDreco}

The neutral $\pizero$ and $\eta$ mesons are reconstructed from photon pairs.
Mass constrained fits are applied to $\pizero$ and $\eta$ candidates.
The photons used to reconstruct the $\eta$ must have energies greater than 
200\,\MeV. 
Any photon used to reconstruct an $\eta$ candidate is vetoed if it
can be paired with an additional photon with energy greater than
$150\, \MeV$ to form a $\pizero$ candidate with mass in the range  
120 - 150\, $\MeV/c^2$. 
This reduces the contribution from the $\Bmi \ra \Dstze \rhomi$ background 
when a photon from a high energy $\pizero$ from $\rho$ decay 
is associated with another photon to form an $\eta$ candidate. 
This veto also reduces the cross-feed from color-suppressed 
$\Dstze \pizero$ modes to the $\Dzero \eta$ channel.

Candidate $\omega$ mesons are reconstructed from $\threepi$ candidates
with a vertex constraint applied to the $\twopi$.
To reduce combinatoric background,
the charged pion candidates must have a momentum greater than 
200\,\MeVc\ while 
the $\pizero$ must have an energy greater than 250\,\MeV.

The $\Dzero$ mesons are reconstructed in three decay modes: 
$\Kpi$, $\Ktwopi$, and $\Kthreepi$.
Vertex constraints are applied to the charged particles and
mass constraints are applied using all particles.
In the $\Kpi$ final state the kaon candidate must satisfy the pion veto 
requirement
while in the $\Ktwopi$ and $\Kthreepi$ final states the kaon candidate must 
satisfy the tight kaon criterion because of the increased background present 
in these combinations.
All pion candidates must fail the tight kaon criterion.
To reduce combinatoric background in 
the $\Ktwopi$ final state, we require in addition 
that either the $\pimi \pizero$ or one of the two $\Kmi \pi$ combinations have 
an invariant mass consistent with an intermediate resonant state 
$\rhomi$ or $\Kstar$.
The energy of the $\pizero$ is also required to be greater than $300\,\MeV$.
 
The $\Bzerobar$ mesons are reconstructed from 
$\Dzero \hzero$ $(\hzero = \pizero, \eta, \omega)$ pairs.
For the final state $\Dzero \omega$ we apply a vertex constraint 
to the $\Dzero$ and the two charged pions.
The energy and momentum of the $\Bzerobar$ are calculated from the
improved energies and momenta of the $\Dzero$ and $\hzero$ that result
from the vertex and mass fits.  

The $\Bzerobar \ra \Dzero \pizero$ sample is contaminated by the
color-allowed $\Bmi \ra \Dzero \rhomi$ decay
that has a branching fraction about fifty times larger than 
the color-suppressed decay.
The contamination is caused by asymmetric $\rhomi \ra \pimi \pizero$ decays
where the $\pizero$ has most of the energy of the $\rhomi$.
The decay channel $\Bmi \ra \Dzero \rhomi$ is reconstructed 
and $\Bzerobar \ra \Dzero \pizero$ candidates that are also reconstructed 
as $\Bmi \ra \Dzero \rhomi$ are vetoed.
To make this veto as efficient as possible, 
the $\rhomi$ is reconstructed using not only pion candidates as defined
above but also low momentum pion candidates that are reconstructed 
using the SVT alone.
The veto reduces the signal efficiency by 10\%.
Other channels vetoed are $\Bzerobar \ra \Dstarzero \hzero$.
$\Dstarzero$ candidates are reconstructed from a $\Dzero$ 
paired with either a $\pizero$ with momentum less than $300\, \MeVc$
in the $\eplemi$ center-of-mass frame or with a photon.
In the latter case, the photon must not be a partner in a pair of photons 
forming a $\pizero$ candidate.
The $\Dstarzero$ -- $\Dzero$ mass difference is required to be
less than 2.0 standard deviations away from its nominal value.

The reconstructed masses of the $\Dzero$, $\pizero$, and $\eta$ 
are required to be within $\pm 2.5 \sigma$ of their nominal 
value \cite{ref:PDG}.
The $\Dzero$ mass resolutions are about 6, 12, and $5 \,\MeV/c^2$ for the  
$\Kpi$, $\Ktwopi$, and $\Kthreepi$ decay modes, respectively,
while the $\pizero$ and
$\eta$ mass resolutions are about 8 and $16\,\MeV/c^2$, respectively.
The reconstructed mass of the $\omega$ candidates is required to be 
within $\pm 25\,\MeV/c^2$ of the $\omega$ nominal value \cite{ref:PDG}.

\subsection{$B$ candidate selection}
\label{subsec:bsel}

Both $b$ and $u, d, s$, and $c$ quark-antiquark production 
contribute combinatoric background events for which the mass
of the candidate $B$ does not peak near the nominal $B$ mass.
To reject the $u, d, s$, and $c$ components, 
we apply several selection criteria based upon the shape of
the event in the $\epl\emi$ center-of-mass frame.

The ratio of the second to the zeroth Fox-Wolfram moment 
\cite{ref:FoxWolf} must be $R_2 < 0.6$. 
For each reconstructed $\Bzerobar$ candidate, 
we compute the thrust and sphericity axes of
both the candidate and the rest of the event \cite{ref:BaBarBook}, 
and we apply a selection
on the angles $\thetathr$ and $\thetasph$ between the two axes, respectively.
The distributions of $|\cos{\thetathr}|$ and $|\cos{\thetasph}|$ peak
near 1.0 for $udsc$ background while they are nearly flat for $B$ decays.
Thus we require $|\cos{\thetasph}| < 0.85$ and $|\cos{\thetathr}| < 0.85$
for the $\Dzero \pizero$ and the $\Dzero \eta$ final states.
For the $\Dzero \eta$ final state we take advantage of $\sin^2{\theta^*}$
distribution of the production
angle $\theta^*$ of the $B$ mesons in the $\eplemi$ center of mass system, 
demanding $|\cos{\theta^*}| < 0.80$.
The corresponding distribution is almost flat for any kind of combinatoric 
background.
For the $\Dzero \omega$ channel, as the $\omega$ is a polarized vector 
particle, we use two other angles.
The first angle is $\theta_N$, 
defined as the angle between the normal to the plane 
of the three daughter pions in
the $\omega$ center-of-mass frame and the $\omega$ direction 
in the $B$ cener-of-mass frame.
The second angle is $\theta_D$, 
the angle between one of the three pions in the $\omega$ 
center-of-mass frame and the direction of one of
the two remaining pions in the center-of-mass frame of these two pions.
The signal events are distributed as 
$\cos^2{\theta_N}$ and $\sin^2{\theta_D}$, 
while the corresponding distributions are
flat for combinatoric background.  
We select only events inside 
a region of the three-dimensional parameter space of the angles $\theta^*$, 
$\theta_N$, and $\theta_D$ with high signal population. 

In a small fraction of the events, even 
after the selection criteria and after the veto of
$\Bmi \ra \Dzero \rhomi$ (for $\Dzero \pizero$) or
$\Bzerobar \ra \Dstarzero \hzero$ (for $\Dzero \eta/\omega$),
more than one $B$ candidate survives.
We select the candidate with the lowest value of
$$\chisq_B = 
\left( \frac{ m_D - m^{\rm nom}_D } 
            {\sigma_{m_D}} \right)^2 + 
\left( \frac{ m_h - m^{\rm nom}_h}
            {\sigma_{m_h}} \right)^2,$$
where
$\sigma_{m_D}$ and $\sigma_{m_h}$ are the average
mass resolutions of the $\Dzero$ and the light hadron $\hzero$.
The $\Dzero$ mass resolution depends on the
$\Dzero$ decay mode.
The ratios in parentheses are found to be approximately Gaussian 
with mean values near 0.0 and standard deviations near 1.0.

\section{Event yields}
\label{sec:Physics}

The energy-substituted mass is defined as 
$\mes = \sqrt{(E_B^*)^2 - (p_B^*)^2}$ and the energy difference is defined as
$\Delta E = E_D^* + E_h^* - E_B^*$.
Here $p_B^*$ is the measured momentum of the $B$ candidate and
$E_D^*$ and $E_h^*$ are the energies of the $\Dzero$ and the $\hzero$, 
all calculated from the measured $\Dzero$ and $\hzero$ momenta. 
$E_B^*$ is the beam energy (and thus the energy of the $B$ meson).
All quantities with a * are calculated in the $\eplemi$ 
center-of-mass frame. 
Signal events have $\mes \approx m_B = 5.279\, \GeVcsq$ and 
$\Delta E \approx 0$.
The $\mes$ resolution is dominated by the beam energy spread and is 
approximately $3\,\MeVcsq$, independent of the $B$ decay mode.
The $\Delta E$ resolutions for the $\Dzero \pizero$ and
$\Dzero \eta$ modes are dominated by the photon energy resolution 
in the EMC for the $\pizero$ or $\eta$ decay products.
They are approximately 30 -- 40\, \MeV.
The $\Delta E$ resolution is better for the $\Dzero \omega$ 
mode because it is dominated by tracking resolution and is
approximately 20\,\MeV.

\begin{figure}
\begin{center}
\scalebox{.5}{\includegraphics{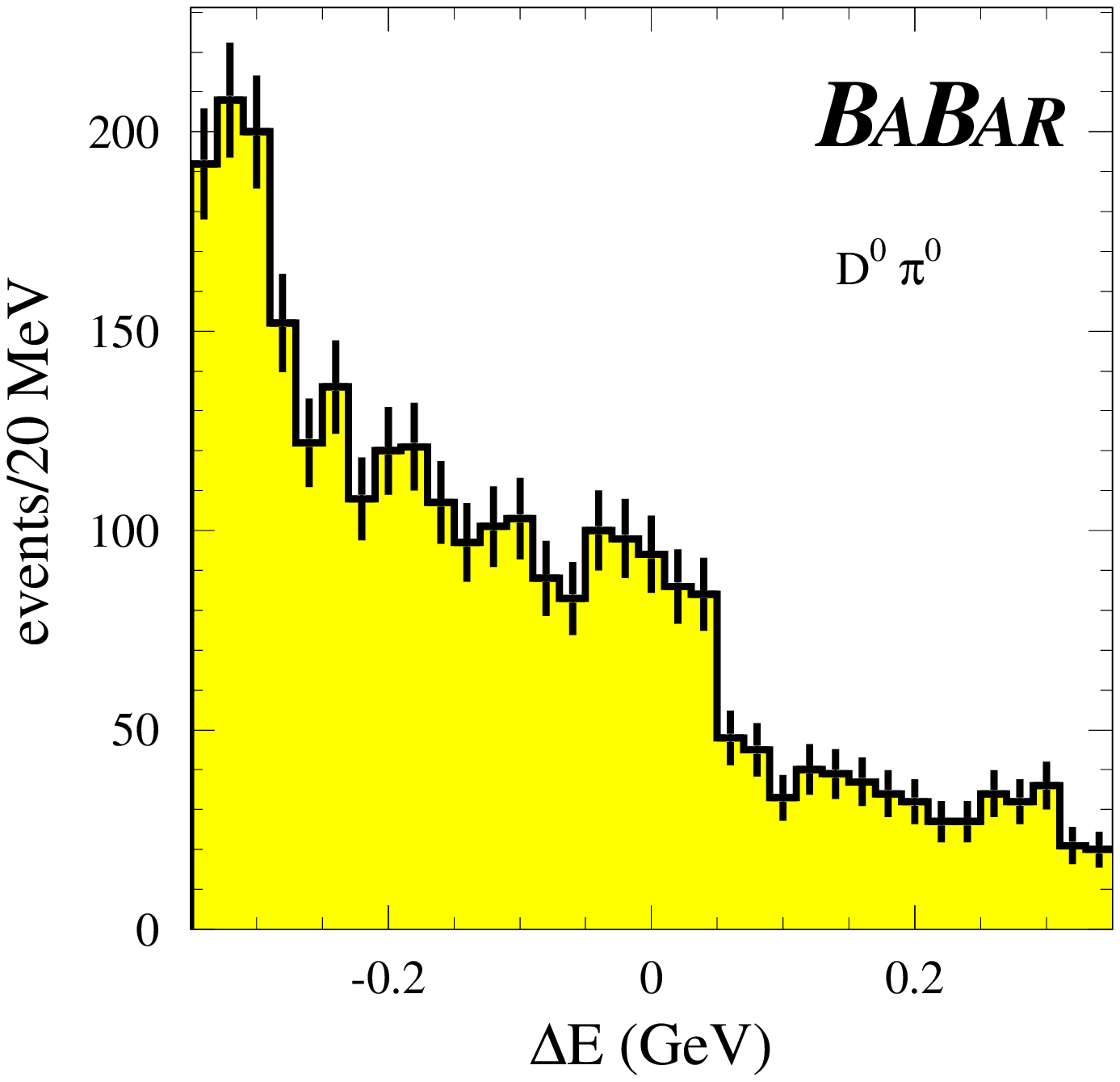}
\includegraphics{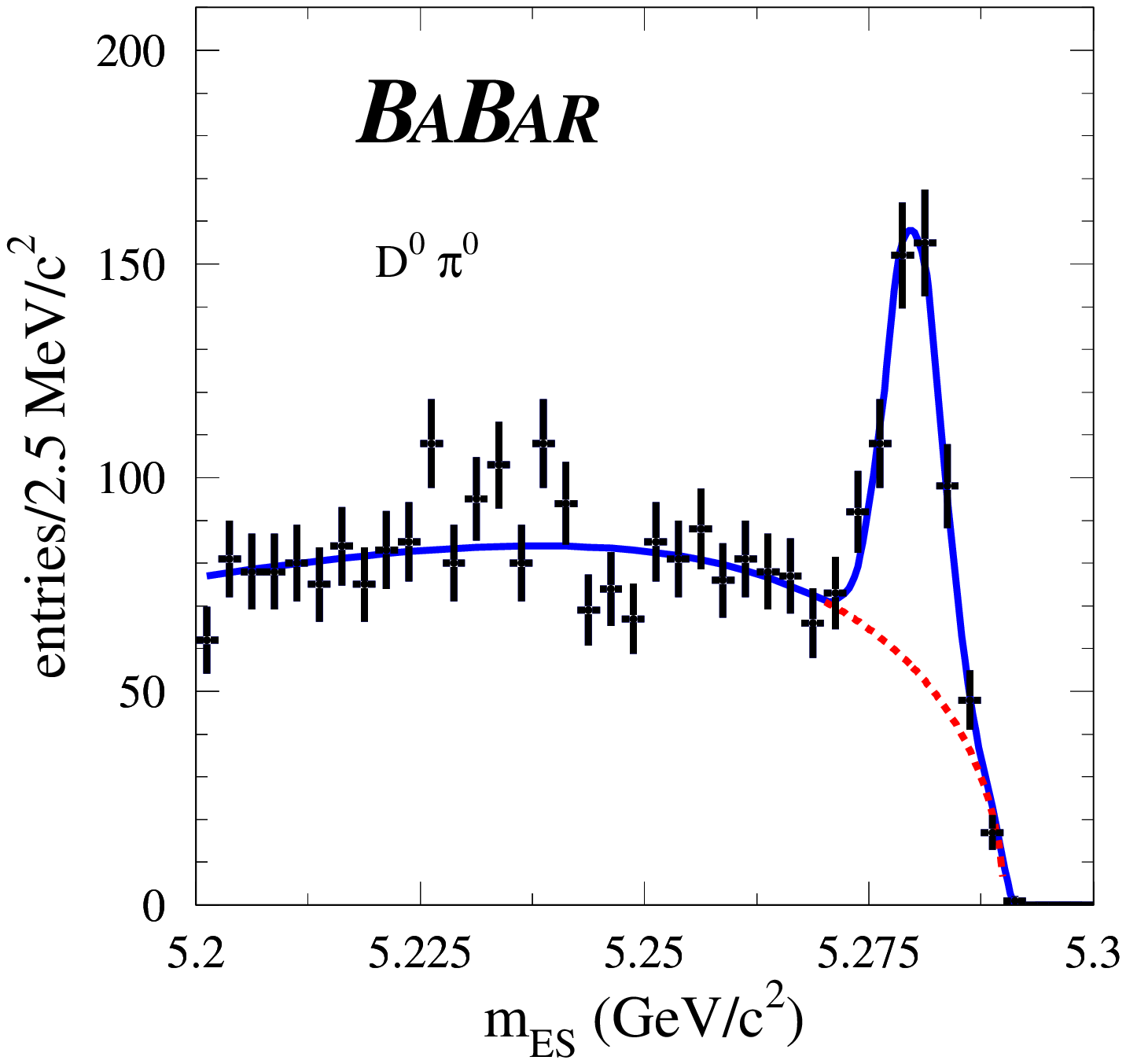}}  
\caption{Distributions of $\Delta \rm{E}$  
(left, with $\mes$ in the range $5.27$--$5.29\,\GeVcsq$)
and beam energy substituted mass $\mes$ 
(right, with $-90 < \Delta E<100\,\MeV$)
for $\Bzerobar \ra \Dzero \pizero$ candidates.}
\label{fig:DEandmESpi0}
\end{center}
\end{figure}

\begin{figure}
\begin{center}
\scalebox{.5}{\includegraphics{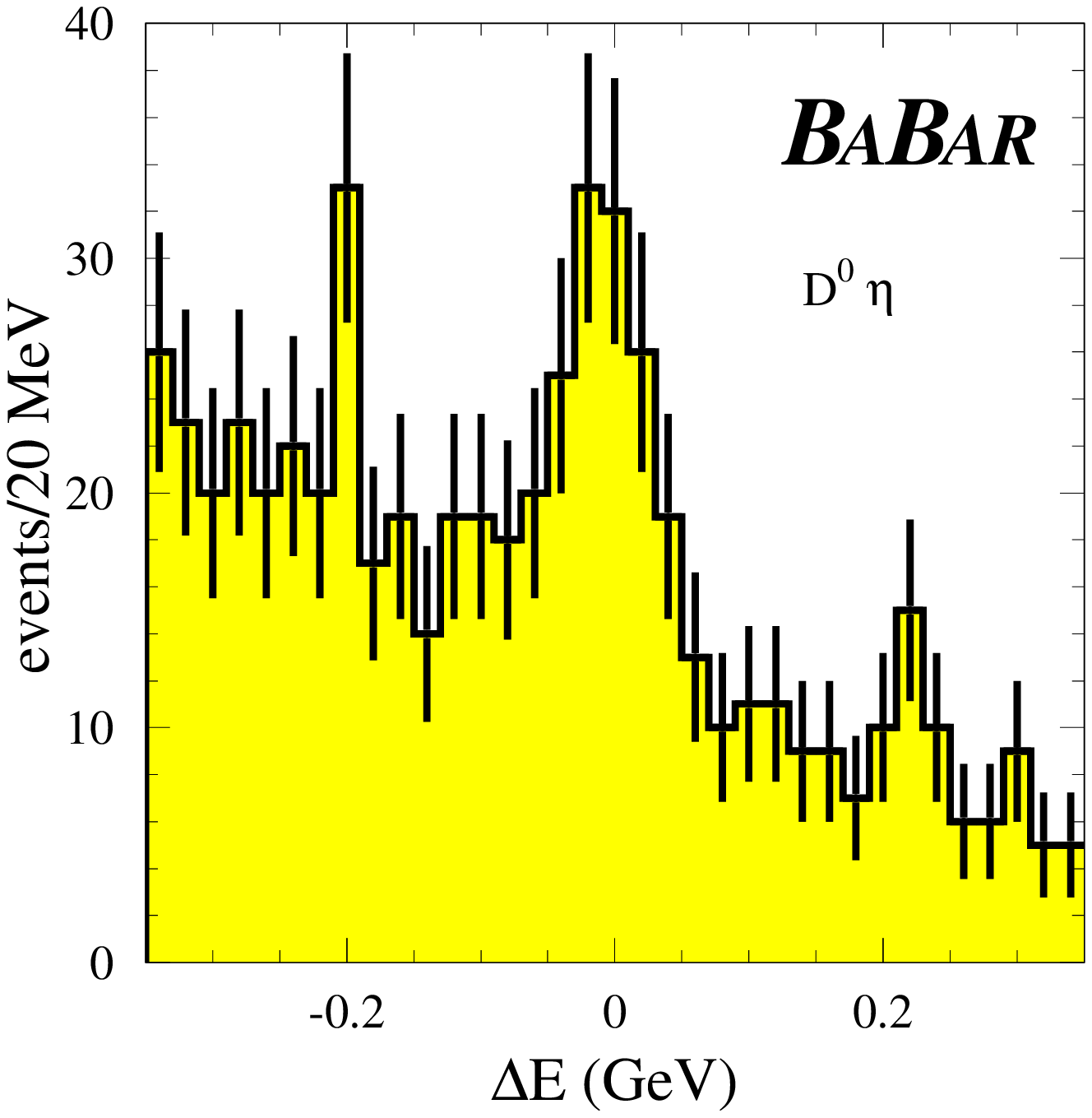}
\includegraphics{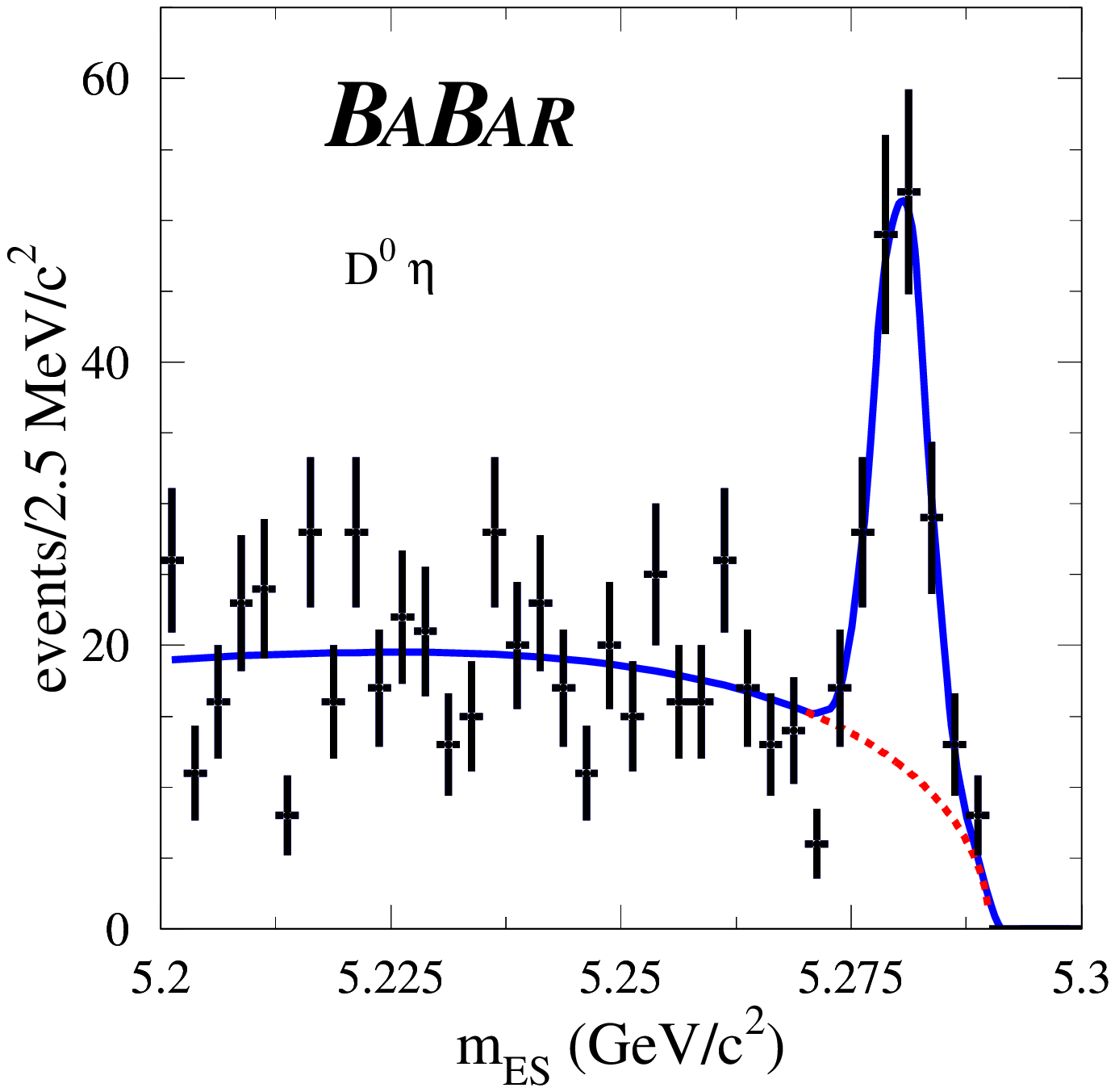}}  
\caption{Distributions of $\Delta \rm{E}$ 
(left, with $\mes$ in the range $5.27$--$5.29\,\GeVcsq$)
and beam energy substituted mass $\mes$ 
(right, with $\vert \Delta E \vert<90\,\MeV$) 
for $\Bzerobar \ra \Dzero \eta$ candidates.}
\label{fig:DEandmESeta}
\end{center}
\end{figure}

\begin{figure}
\begin{center}
\scalebox{.5}{\includegraphics{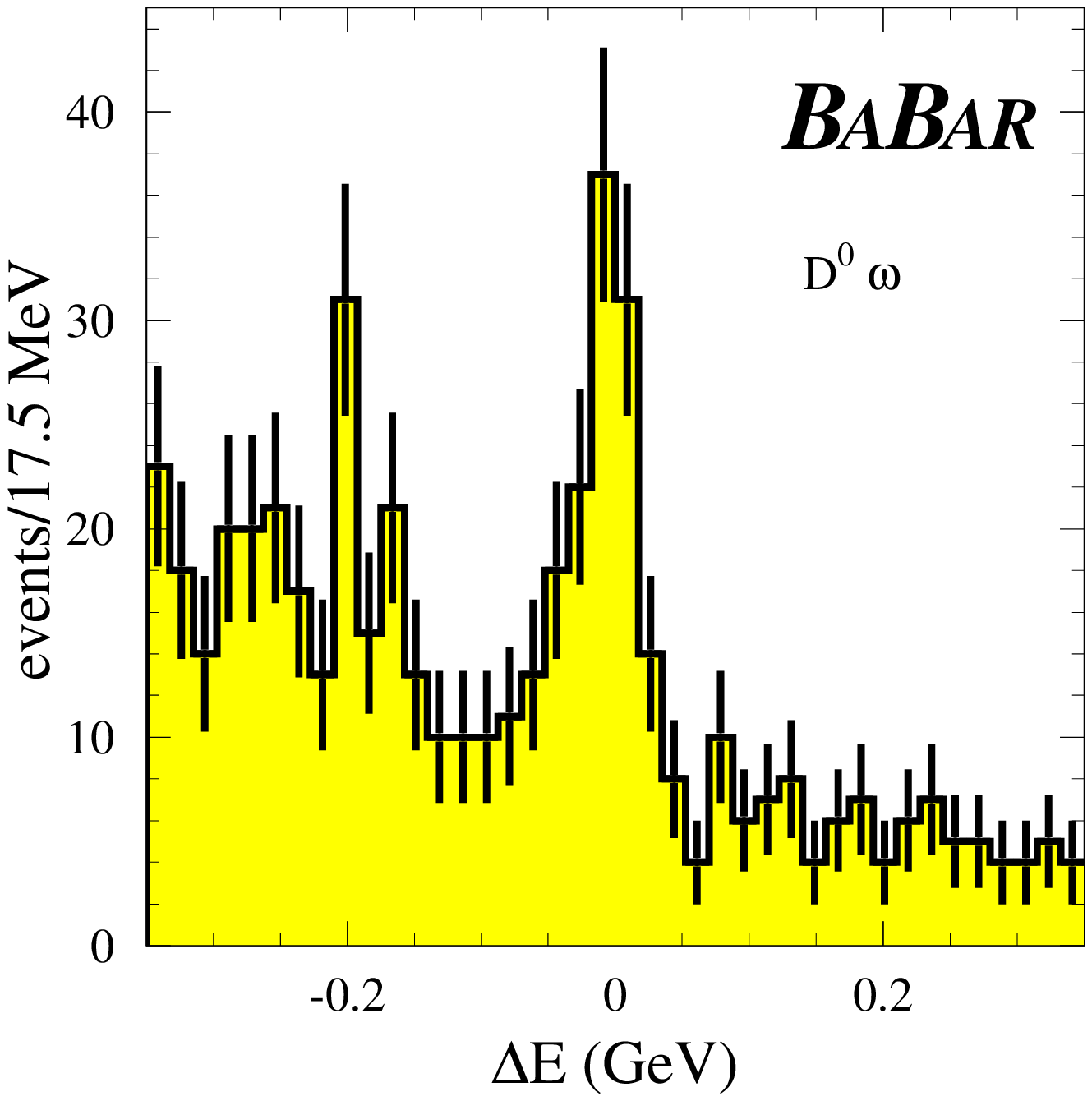}
\includegraphics{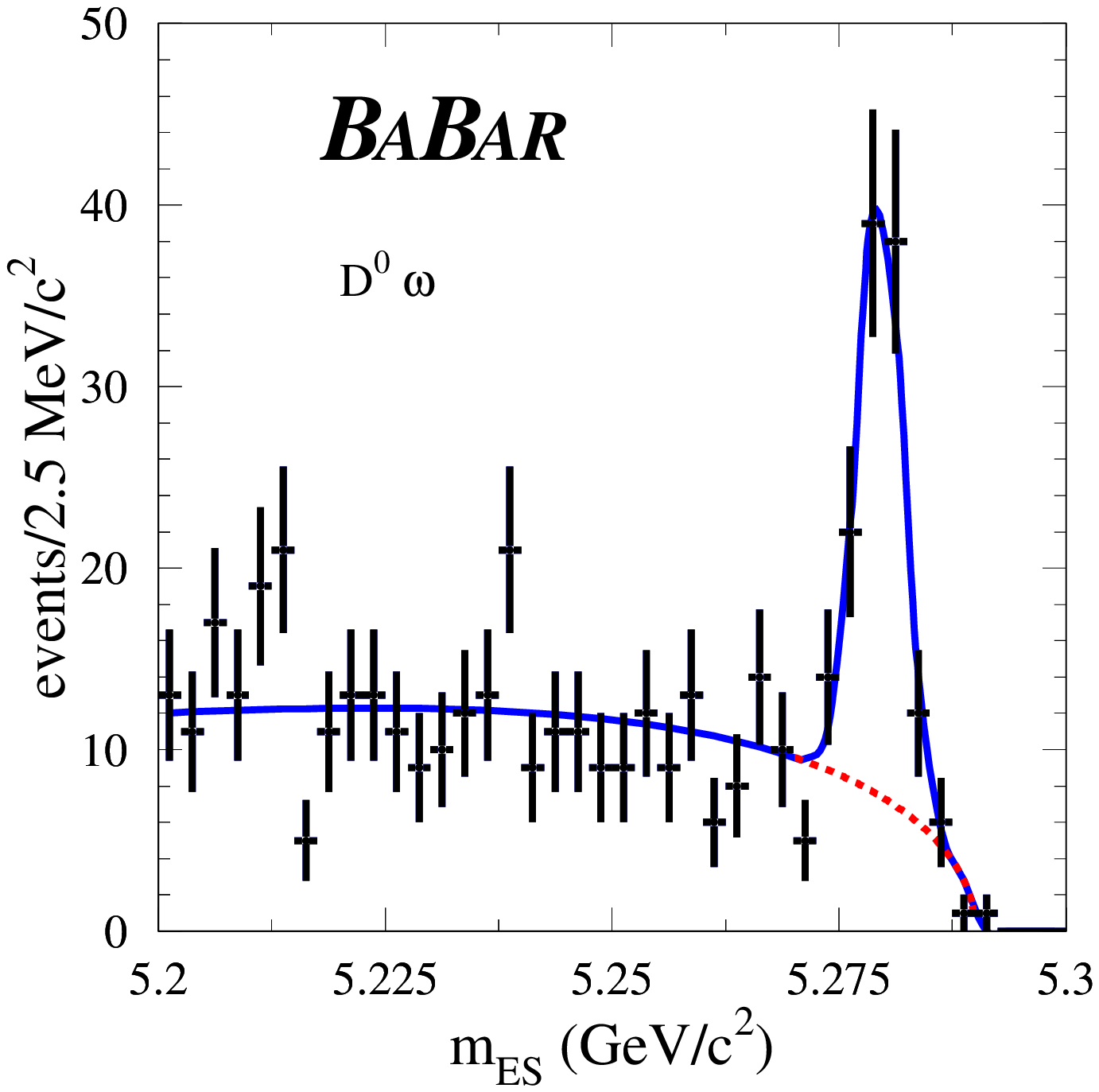}}  
\caption{Distributions of $\Delta \rm{E}$ 
(left, with $\mes$ in the range $5.27$--$5.29\,\GeVcsq$)
and beam energy substituted mass $\mes$ 
(right, with $\vert \Delta E \vert<60\,\MeV$)
for $\Bzerobar \ra \Dzero \omega$ candidates.}
\label{fig:DEandmESomega}
\end{center}
\end{figure}

The $\mes$ distributions with a selection on $\Delta E$ 
and the $\Delta E$ distributions with a selection on $\mes$ 
for $\Dzero \pizero$, $\Dzero \eta$, and $\Dzero \omega$ are shown in
Figs.~\ref{fig:DEandmESpi0}, \ref{fig:DEandmESeta}, 
and \ref{fig:DEandmESomega}, respectively.
The $\mes$ distributions are shown for 
$-90 < \Delta E < 100\,\MeV$ for $\Dzero \pizero$,
$\vert \Delta E \vert < 90\,\MeV$ for $\Dzero \eta$, and
$\vert \Delta E \vert < 60\,\MeV$ for the $\Dzero \omega$ final states,
respectively.
The $\Delta E$ distributions are shown for $\mes$ 
in the range $5.27$--$5.29\,\GeVcsq$.
We observe clear signals in all three channels.

We perform a least-squares fit of a function consisting of the sum of 
a Gaussian and an Argus background function \cite{ref:Argus} 
to the $\mes$ distribution for the $\Dzero \pizero$ final state.
For the $\Dzero \eta$ and $\Dzero \omega$ modes,
where the signal yields are lower, we perform an unbinned maximum
likelihood fit of the same functions to the $\mes$ distribution.
The Argus function accounts for random combinatoric background
originating from both $udsc$ continuum, $\tau$ leptons, two-photon processes,
and $B \Bbar$ events but not for ``peaking background''. The peaking 
background accounts for specific channels described by a Gaussian 
very similar to the real signal.

We investigate peaking background by studying a generic
background $B \Bbar$ \mc\ sample.
The only significant peaking background in the $\Bzerobar \ra \Dzero \pizero$ 
selection originates from
$\Bmi \ra \Dstze \rhomi$ with an undetected low-momentum $\pimi$. 
According to the \mc\ simulation, 
the veto described in Sec.~\ref{subsec:X0andDreco} 
rejects $75\%$ of $\Bmi \ra \Dstze \rhomi$ events in the range
$\Delta E < -90\, \MeV$ but only 16\% of that background
in the signal range $-90 < \Delta E < 100\, \MeV$.
The veto thus causes a flattening of the background 
in the $\Delta E$ distribution thereby reducing the systematic error
from the uncertainty in the energy resolution of the EMC.
The peaking background is a smaller problem for $\Dzero \eta$ and 
$\Dzero \omega$ modes and contributes less than 12\% 
of the total background in the signal region.
We use the generic $B \Bbar$ \mc\ sample
to estimate the amount of peaking background.
Uncertainty in the branching fraction is included as a systematic error.

For the $\Dzero \pizero$ final state, the least-squares fit 
to the \mes\ distribution in Fig.\ref{fig:DEandmESpi0}
gives the uncorrected event yield 
as the area under the Gaussian. 
The $\Dzero \eta$ and $\Dzero \omega$ modes have smaller signal yields
so the contribution parametrized by the Argus function is subtracted 
from the event yield obtained from the number of entries 
in the signal region defined by $5.27 < \mes < 5.29$ and 
$\vert \Delta E \vert < 90\,\MeV$ for $\Dzero \eta$ and
$\vert \Delta E \vert < 60\,\MeV$ for the $\Dzero \omega$ final states.
Finally, after subtraction of the estimated peaking background 
we obtain the signal yield.
We obtain $291 \pm 31$, $101 \pm 14$, and $78 \pm 12$ for the
$\Dzero \pizero$, $\Dzero \eta$ and $\Dzero \omega$ final states,
respectively, see Table \ref{tab:BF}.

Event yields must be corrected for cross-feed from other color 
suppressed modes.
Cross-feed to each signal from 
$\Bzerobar \ra \Dstze \hzero$ decays is investigated using 
the branching fractions measured by the 
CLEO \cite{ref:CLEO} and Belle \cite{ref:Belle} Collaborations 
for the $\Dstarzero \hzero$ final states. 
For each $\hzero$ the dominant 
contribution to $\Bzerobar \ra \Dzero \hzero$ arises from the associated 
$\Bzero \ra \Dstarzero \hzero$ modes. 
In the signal region, we estimate that the event yields for 
the $\Dzero \pizero$, $\Dzero \eta$, and $\Dzero \omega$ 
final states receive contributions equal to 4.5\%, 8.3\%, and 2.4\% from
cross-feed.

The acceptance $\cal{A}$, as determined from signal \mc\ samples,
must be corrected for differences between data and \mc\
simulation in tracking, vertex fitting, and particle identification.
We correct the \mc\ simulation results using the outcome of
detailed studies of detector performances
in which control sets of data are compared with their \mc\ simulation.
These procedures provide corrections
that are applied per track (for track reconstruction efficiency), 
per kaon candidate (for particle identification efficiency),
and per vertex fit (for vertex fit efficiency).

Dividing corrected signal yields (${\cal S}$) by 
the number of $B\Bbar$ events in the data sample ${N(\BBbar)}$, 
the corrected acceptances (${\cal A}$), 
and the secondary branching fractions 
of the $\Dzero$ and the $\hzero$ 
into the reconstructed final states $X$ and $Y$ respectively,
gives branching fractions as
$${\cal B}(\Bzerobar \ra \Dzero \hzero) = 
\frac {\cal S} { {N(\BBbar)} \times {\cal A \times B}(\Dzero\ra X) \times 
{\cal B}(\hzero\ra Y) }.$$
The resulting branching fractions and their statistical errors are listed 
in Table \ref{tab:BF}.

\section{Systematic uncertainties and results}
\label{sec:EffSystematics}

Systematic errors are associated with the corrections discussed above. 
In addition, we have considered systematic errors from other sources.
Uncertainties in the acceptances 
from photon detection
account for imperfect simulation of photon energy and position resolution,
thus affecting $\pizero$ and $\eta$ reconstruction and
the $\Delta \rm{E}$ resolution.
We have also investigated uncertainties in the simulation of peaking and 
combinatoric background. 
For the $\Dzero \pizero$ mode
the systematic error associated with the veto of the $\Bmi \ra \Dzero \rhomi$ 
background has been studied and is part of the systematic error
on the  background estimate.
We have varied the selection criteria described in Sec.~\ref{subsec:partsel} and 
\ref{subsec:bsel} in order to assign 
a systematic error to the event selection. 
The errors from the counting of $B \Bbar$ pairs,
from the branching fractions of $\Dzero$ and $\hzero$ secondary decays
\cite{ref:PDG}, and the statistical error from 
the \mc\ samples used to determine the signal acceptance 
have also been evaluated.   
Table~\ref{tab:syst} 
summarizes these systematic errors for the three 
final states $\Dzero \pizero$, $\Dzero \eta$, and  $\Dzero \omega$.

\def\AB{${\cal A} \times {\cal B}(\Dzero \ra X) \times
{\cal B}(\hzero \ra Y)$}

\begin{table}[!htb]
\caption{Signal event yields, \AB, and preliminary 
${\cal B}(\Bzerobar \ra \Dzero \hzero)$.
The signal event yields shown are not corrected for cross-feed.}

\begin{center}
\begin{tabular}{|c|c|c|c|}
\hline
$\Bzerobar$ decay & Event yield  & \AB (\%) & ${\cal B}(\Bzerobar \ra \Dzero \hzero) (10^{-4})$    \\
\hline
$\Dzero \pizero$  & $291 \pm 31$ & 2.1      & $2.89 \pm 0.29 (\stat) \pm 0.38 (\syst)$ \\
$\Dzero \eta$     & $101 \pm 14$ & 0.9      & $2.41 \pm 0.39 (\stat) \pm 0.32 (\syst)$ \\
$\Dzero \omega$   &  $78 \pm 12$ & 0.6      & $2.48 \pm 0.40 (\stat) \pm 0.32 (\syst)$ \\  
\hline
\end{tabular}
\end{center}
\label{tab:BF}
\end{table}

\begin{table}[!htb]
\caption{Fractional systematic errors on the measured branching fractions.}
\begin{center}
\begin{tabular}{|c|c|c|c|} \hline
Category                                    & $\Dzero \pizero$ (\%)  & $\Dzero \eta$ (\%)& $\Dzero \omega$ (\%) \\
\hline
Tracking                                    & 2.1                    & 2.0               & 3.6                  \\
Vertex fit                                  & 1.4                    & 1.4               & 2.5                  \\
Kaon identification                         & 2.5                    & 2.5               & 2.5                  \\
Cross feed                                  & 2.3                    & 4.3               & 1.2                  \\
$\gamma$, $\pizero$, and $\eta$ detection   & 5.3                    & 3.6               & 6.0                  \\
$\Delta E$ resolution                       & 5.7                    & 6.7               & 4.6                  \\
Background estimate                         & 4.4                    & 3.2               & 5.2                  \\
Event selection                             & 7.8                    & 7.6               & 5.3                  \\
Number of $B\Bbar$ pairs                    & 1.1                    & 1.1               & 1.1                  \\
${\cal B}(\Dzero)$ and ${\cal B}($\hzero$)$ & 4.2                    & 4.5               & 4.5                  \\
\mc\ statistics                             & 0.7                    & 1.5               & 2.4                  \\
\hline
Total                                       & 13.3                   & 13.5              & 12.9                 \\
\hline
\end{tabular}
\end{center}
\label{tab:syst}
\end{table}

We obtain the branching fractions 
${\cal B} (\Bzerobar \ra \Dzero \pizero) = 
(2.89 \pm 0.29 (\stat) \pm 0.38 (\syst)) \times 10^{-4}$,
${\cal B} (\Bzerobar \ra \Dzero \eta) = 
(2.41 \pm 0.39 (\stat) \pm 0.32 (\syst)) \times 10^{-4}$, 
and ${\cal B} (\Bzerobar \ra \Dzero \omega) = 
(2.48 \pm 0.40 (\stat) \pm 0.32 (\syst)) \times 10^{-4}$.
They are listed in Table \ref{tab:BF}.
These results are preliminary.
The branching fractions are in good agreement with previous results from
the CLEO \cite{ref:CLEO} and Belle \cite{ref:Belle} Collaborations. 
They are more precise mainly due to larger sample of B decays.
The branching fractions are considerably larger than the factorization predictions
for these three modes.


\section{Acknowledgments}
\label{sec:Acknowledgments}

We are grateful for the 
extraordinary contributions of our \pep2\ colleagues in
achieving the excellent luminosity and machine conditions
that have made this work possible.
The success of this project also relies critically on the 
expertise and dedication of the computing organizations that 
support \babar.
The collaborating institutions wish to thank 
SLAC for its support and the kind hospitality extended to them. 
This work is supported by the
US Department of Energy
and National Science Foundation, the
Natural Sciences and Engineering Research Council (Canada),
Institute of High Energy Physics (China), the
Commissariat \`a l'Energie Atomique and
Institut National de Physique Nucl\'eaire et de Physique des Particules
(France), the
Bundesministerium f\"ur Bildung und Forschung and
Deutsche Forschungsgemeinschaft
(Germany), the
Istituto Nazionale di Fisica Nucleare (Italy),
the Research Council of Norway, the
Ministry of Science and Technology of the Russian Federation, and the
Particle Physics and Astronomy Research Council (United Kingdom). 
Individuals have received support from 
the A. P. Sloan Foundation, 
the Research Corporation,
and the Alexander von Humboldt Foundation.

\end{document}